\pdfoutput=1

\documentclass[11pt,twoside,a4paper,cmspaper,final,collab]{cms-tdr}
\def\svnVersion{262098}\def\cmsCernNo{2013-037}\def\cmsCernDate{\today}\def\cmsMessage{Published in the Journal of High Energy Physics as \href{http://dx.doi.org/10.1007/JHEP09(2014)094}{\doi{10.1007/JHEP09(2014)094.}}}

\begin{document}\cmsNoteHeader{BPH-11-021}

\hyphenation{had-ron-i-za-tion}
\hyphenation{cal-or-i-me-ter}
\hyphenation{de-vices}
\RCS$HeadURL: svn+ssh://svn.cern.ch/reps/tdr2/papers/BPH-11-021/trunk/BPH-11-021.tex $
\RCS$Id: BPH-11-021.tex 262065 2014-09-26 14:46:36Z spanier $

\cmsNoteHeader{BPH-11-021}
\title{Measurement of prompt $\JPsi$ pair production in pp collisions at $\sqrt{s}= 7\TeV$}

\date{\today}

\abstract{
Production of prompt $\JPsi$
meson pairs in proton-proton collisions at $\sqrt{s}=7$\TeV is measured
with the CMS experiment at the LHC in a data sample corresponding
to an integrated luminosity of about 4.7\fbinv.
The two $\JPsi$ mesons are fully
reconstructed via their decays into $\mu^+ \mu^-$ pairs.
This observation provides for the first time access to the high-transverse-momentum
region of $\JPsi$
pair production where model predictions are not yet established.
The total and differential cross sections are measured in a phase space
defined by the individual
$\JPsi$ transverse
momentum ($\pt^{\JPsi}$) and
rapidity ($\abs{y^{\JPsi}}$):
$\abs{y^{\JPsi}}<1.2$ for
$\pt^{\JPsi}>6.5\GeVc$;
$1.2<\abs{y^{\JPsi}}<1.43$ for a
$\pt$ threshold that scales linearly with
$\abs{y^{\JPsi}}$ from 6.5 to 4.5\GeVc; and
$1.43<\abs{y^{\JPsi}}<2.2$ for
$\pt^{\JPsi}>4.5\GeVc$.
The total cross section, assuming unpolarized prompt
$\JPsi$ pair production is
$1.49 \pm 0.07\stat \pm 0.13\syst$\unit{nb}.
Different assumptions about the
$\JPsi$ polarization
imply modifications to the cross section ranging from $-$31\%
to $+$27\%.
}

\hypersetup{
pdfauthor={CMS Collaboration},
pdftitle={Measurement of prompt J/psi pair production in pp
collisions at sqrt(s) = 7 TeV},
pdfsubject={CMS},
pdfkeywords={CMS, B physics}}

\maketitle

\section{Introduction}

The measurement of \JPsi meson pairs that are directly created in the primary
interaction (prompt) in proton-proton (pp) collisions at $\sqrt{s}=7$\TeV
provides general insight into how particles are produced during proton
collisions at the CERN LHC.
Owing to the high flux of incoming partons at the LHC energy,
it is expected that more than one parton pair will often scatter in
a pp collision~\cite{Kom:2011bd}.  These multiparton scattering contributions
are difficult to address within the framework of perturbative quantum
chromodynamics (QCD), hence the need for experimental studies (see \eg,
Ref.~\cite{Ko:2010xy} and references therein).  The general assumption is
that single-parton scattering (SPS) is the dominant process.  Double-parton
scattering (DPS) and higher-order multiple-parton interactions are widely
invoked to account for observations that cannot be explained otherwise, such
as the rates for multiple heavy-flavor production~\cite{Berger:2009cm}.
New measurements will help the creation of
more realistic particle production models. The production of
\JPsi meson pairs provides a clean signal in a parton-parton interaction
regime that is complementary to the one probed by studies based on hadronic
jets.  Multiple-parton interactions can lead to distinct differences in
event variables that probe pair-wise balancing, such as the absolute
rapidity difference $\abs{\Delta y}$ between the two \JPsi mesons~\cite{Baranov:2011ch,Baranov:2013,Kom:2011bd}.  The strong correlation
of two \JPsi mesons produced via SPS interaction results in small values of
$\abs{\Delta y}$, whereas large values of $\abs{\Delta y}$ are possible for
production due to DPS.

In contrast to earlier experiments where quark-antiquark annihilation
dominated~\cite{Badier:1982ae,Badier:1985ri}, the dominant \JPsi production
process in pp collisions at the LHC is gluon-gluon
fusion~\cite{Humpert:1983yj}.  At the parton level, the two \JPsi mesons
are either produced as color-singlet states or color-octet states that turn
into singlets after emitting gluons.  Color-octet contributions for \JPsi pair production at transverse momentum (\pt) of a pair below 15\GeVc and low
invariant mass are considered to be negligible, but play a greater role as
\pt increases~\cite{Berezhnoy:2011xy,Qiao:2009kg}.  Next-to-leading-order
QCD calculations also indicate enhanced contributions from color-singlet
heavy-quark pair production at higher \pt \cite{Campbell:2007ws,Artoisenet:2007xi,Gong:2008hk,PRL.111.122001}.
The CMS experiment provides access to \pt measurements above 15\GeVc.

Recently, the LHCb experiment measured the cross section for \JPsi pair
production in pp collisions at $\sqrt{s}=7$\TeV
to be $5.1 \pm 1.0 \pm 1.1\unit{nb}$ (where the first uncertainty is
statistical and the second systematic) within the LHCb phase space (defined
as $2 <y^{\JPsi}<4.5$ and $\pt^{\JPsi}<10\GeVc$)~\cite{Aaij:2012dz}.
Theoretical calculations of \JPsi pair production via SPS based on
leading-order color-singlet states predict a cross section of 4\unit{nb},
with an uncertainty of about 30\%~\cite{Berezhnoy:2011xy,Berezhnoy:2012xq}.
This prediction is consistent with
the measured value. The CMS experiment samples a \JPsi production regime
complementary to LHCb, with coverage at higher \pt and more central rapidity.
Hence, \JPsi pair production cross section measurements by CMS provide
new information for the development of production models that include
higher-order corrections and DPS.

Model descriptions of \JPsi pair production are also a crucial input
to quantify nonresonant contributions in the search for resonances.
States can be searched for
with CMS in a wider \JPsi pair invariant-mass range as
compared to previous experiments. For example, the bottomonium ground state
$\eta_b$ is expected to decay into two \JPsi mesons in analogy to the
$\eta_c$ charmonium ground state that decays into two $\phi$
mesons~\cite{Maltoni:2004hv}.  However, explicit calculations based on
nonrelativistic QCD (NRQCD)~\cite{Braaten:2000cm,Jia:2006rx} predict
this decay mode to be highly suppressed, so any observation of this
process could indicate possible shortcomings of present NRQCD approaches.
Other predicted resonant states that could decay into two \JPsi mesons
are exotic tetraquark charm states~\cite{Berezhnoy:2011xy}.  A CP-odd
Higgs boson, \eg, in the next-to-minimal supersymmetric standard
model~\cite{Dermisek:2005ar}, is predicted with a mass near
the $\eta_b$.  Mixing with a CP-odd Higgs boson could alter
the behavior of the $\eta_b$ with respect to QCD
predictions~\cite{Domingo:2009tb,Domingo:2010am}.  The BaBar experiment first
observed the $\eta_b$ state in radiative $\Upsilon$
transitions~\cite{Aubert:2009as} and published an upper limit on the
effective coupling of a CP-odd Higgs boson with mass below
9.3\GeVcc to b quarks~\cite{babarhiggs}.
No evidence for a CP-odd Higgs boson was
found by CMS in the $\mu^+\mu^-$ invariant-mass spectrum for masses
between 5.5 and 14\GeVcc~\cite{Chatrchyan:2012am}.

This Letter presents a measurement of the cross section for prompt
\JPsi pair production with data recorded with the CMS experiment
in pp collisions at a center-of-mass energy of 7\TeV.
Acceptance corrections are calculated based on the measured
\JPsi meson kinematics, and efficiency corrections are calculated based
on the measured decay-muon kinematics of each event thereby minimizing the
dependence on production models. Monte Carlo (MC) simulation
samples for different production models with either strongly correlated
\JPsi mesons (SPS model) or less correlated \JPsi mesons (DPS model)
are only used to define the phase-space region and validate the
correction method.  They also provide guidance for the parameterization
of various kinematic distributions in the events.  The SPS generator is a
color-singlet model~\cite{Berezhnoy:2011xy} implemented in
{\PYTHIA6}~\cite{Sjostrand:2006za}, and the DPS generator is implemented
in {\PYTHIA8}~\cite{Sjostrand:2007gs} using color-singlet and -octet
production models.

The cross section measurement is evaluated in a predefined region of the
\JPsi phase space that, in turn, is constrained by the muon
identification and
reconstruction capabilities of CMS.  The differential cross section of
\JPsi pair production is calculated as
\begin{linenomath}
\begin{equation}
\frac{\rd\sigma ({\Pp\Pp}\to \JPsi\,\JPsi+X)}{\rd{}x} =\\ \sum_i
\frac{s_i} { a_i \cdot \epsilon_i
\cdot (BF)^2\cdot \Delta x
\cdot \mathcal{L}}.
\end{equation}
\label{eq:cross-section}
\end{linenomath}
The sum is performed over events $i$ in an interval $\Delta x$, where $x$
represents a kinematic variable describing the \JPsi pair.
In this analysis, $x$ is taken as the invariant mass of the \JPsi pair
($M_{\JPsi\,\JPsi}$), the absolute difference in \JPsi meson rapidities
($\abs{\Delta y}$), and the transverse momentum of the \JPsi pair
($\pt^{\JPsi\,\JPsi}$).  The quantity $s_i$ is the signal weight per event.  The
acceptance value $a_i$ calculated for each event represents the
probability that the muons resulting from the \JPsi decays pass the muon
acceptance.  The detection efficiency $\epsilon_i$ is the
probability for the four muons in an event to be detected and pass the
trigger and
reconstruction quality requirements.  The integrated luminosity of
the dataset is $\mathcal{L}$, and $BF$ is the branching fraction for the \JPsi decay into two muons.
The total cross section in the \JPsi phase-space window is determined by
summing over all events.

\section{CMS detector}
\label{sec:cms_detector}

A detailed description of the CMS detector can be found
elsewhere~\cite{Chatrchyan:2008zzk}.  The primary components used in
this analysis are the silicon tracker and the muon systems.
The tracker operates in a 3.8\unit{T} axial magnetic field generated by a
superconducting solenoid with an internal diameter of 6\unit{m}.  The
innermost part of the tracker consists of three cylindrical layers of pixel
detectors complemented by two disks in the forward and backward directions.
The radial region between 20 and 116\cm is occupied by several layers
of silicon strip detectors in barrel and disk configurations.  Multiple
overlapping layers ensure a sufficient number of hits to precisely
reconstruct tracks in the pseudorapidity range $\abs{\eta} < 2.4$, where
$\eta = -\ln{[\tan{(\theta/2)}]}$\ and $\theta$\ is the polar angle of
the track measured from the positive $z$ axis.  The coordinate
system is defined to have its origin at the center of the detector,
the $x$ axis pointing to the center of the LHC ring, the $y$ axis pointing
up (perpendicular to the plane of the LHC ring), and the $z$ axis aligned with
the counterclockwise-beam direction.  An impact parameter resolution around
15\mum and a \pt resolution around 1.5\% are achieved for charged
particles with \pt up to 100\GeVc.  Muons are identified in the
range $\abs{\eta}< 2.4$, with detection planes made of drift tubes, cathode
strip chambers, and resistive-plate chambers embedded in the steel
flux-return yoke of the solenoid.  The CMS detector response is
determined with MC simulations using \GEANTfour~\cite{Agostinelli2003250}.

\section{Event selection and efficiencies}
\label{sec:reco}
This analysis uses an unprescaled muon trigger path designed to achieve
the highest possible signal-to-noise ratio and efficiency for \JPsi pair
searches during the 2011 data taking.
This trigger requires the presence of at least three
muons, two of which must be oppositely charged, have a dimuon invariant
mass in the interval between 2.8 and 3.35\GeVcc, and a vertex fit
probability greater than 0.5\%, as determined by a Kalman filter
algorithm~\cite{Fruhwirth:1987fm}.  Reconstruction of muons proceeds by
associating measurements in the muon detectors with tracks found in the
silicon tracker, both called segments.  A given muon segment can be
associated with more than one silicon track at the time of reconstruction,
allowing reconstructed muons to share segments in the muon system.
An arbitration algorithm then assigns each muon segment to a unique
muon track.  Muons are further required to pass the following
quality criteria: (i) the associated track segment must have hits in at least
two layers of the pixel tracker and at least 11 total silicon tracker hits
(pixel and strip detectors combined), and (ii) the silicon track fit $\chi^2$ divided by the number of degrees of freedom must be less than 1.8. Three of the muons are required to fulfill the criteria
\begin{linenomath}
\begin{equation}
\begin{aligned}
 \label{eq:tpselect}
  &\pt^\mu > 3.5\GeVc &\qquad\text{if}\quad && \abs{\eta^\mu}<1.2,\\
  &\pt^\mu > 3.5 \rightarrow 2\GeVc  &\qquad\text{if}\quad && 1.2<\abs{\eta^\mu}<1.6, \\
  &\pt^\mu > 2\GeVc  &\qquad\text{if}\quad && 1.6<\abs{\eta^\mu}<2.4,
\end{aligned}
\end{equation}
\end{linenomath}
where the $\pt$\ threshold scales linearly downward with
$\abs{\eta^\mu}$ in the range
$1.2<\abs{\eta^{\mu}}<1.6$.  They must further be matched to the muon
candidates that triggered the event. The fourth muon (not required to match to
the trigger muon candidates) is allowed to pass the looser acceptance criteria
\begin{linenomath}
\begin{equation}
\begin{aligned}
 \label{eq:looseselect}
  &\pt^\mu > 3\GeVc &\qquad\text{if}\quad &&\abs{\eta^\mu}<1.2,\\
  & p^\mu > 3\GeVc &\qquad\text{if}\quad  && 1.2<\abs{\eta^\mu}<2.4,
\end{aligned}
\end{equation}
\end{linenomath}
where $p^\mu$ is the magnitude of the total muon momentum.

Candidate events must have two pairs of opposite-sign muons each with an invariant mass close to the \JPsi mass~\cite{PDG2012}.
Each \JPsi candidate is further required to be within the phase space
\begin{linenomath}
\begin{equation}
\begin{aligned}
 \label{eq:psiselect}
  &\pt^{\JPsi} > 6.5\GeVc &\qquad\text{if}\quad &&\abs{y^{\JPsi}}<1.2,\\
  &\pt^{\JPsi} > 6.5 \rightarrow 4.5\GeVc  &\qquad\text{if}\quad && 1.2<\abs{y^{\JPsi}}<1.43, \\
  &\pt^{\JPsi} > 4.5\GeVc  &\qquad\text{if}\quad && 1.43<\abs{y^{\JPsi}}<2.2,
\end{aligned}
\end{equation}
\end{linenomath}
where the $\pt^{\JPsi}$\ threshold scales linearly with $\abs{y^{\JPsi}}$
in the range $1.2<\abs{y^{\JPsi}}<1.43$. The boundaries are optimized to obtain maximum coverage of the \JPsi phase space within the muon acceptance. If there are more than two \JPsi candidates in an event,
the candidates with the highest vertex fit probabilities are
selected.  For signal MC simulation samples in which multiple
collision events per bunch crossing (pileup events) are included,
this selection process finds the correct dimuon combinations for
99.7\% of the selected events.

In addition to the invariant mass of each dimuon candidate, $m^{\JPsi}$,
two event variables sensitive to the prompt \JPsi pair topology are defined:
(i) the proper transverse decay length, $ct_{xy}$, of
the higher-$\pt$ \JPsi, and (ii) the separation significance, $\delta d$,
between the \JPsi mesons.  Calculating the proper transverse decay
length requires identification of the primary vertex in an event,
defined as the vertex formed by charged-particle tracks with the
highest sum of \pt squared that can be fit to a common
position, excluding the muon tracks from the two
\JPsi candidates.  The transverse decay length in the laboratory frame is
given as $L_{xy} = (\vec{r}_\mathrm{T} \cdot \ptvec^{\JPsi})/\pt^{\JPsi}$,
where $\vec{r}_\mathrm{T}$ is the vector pointing from the primary vertex
to the \JPsi vertex in the transverse plane.
The proper transverse decay length is then
calculated as $ct_{xy} = (m^{\JPsi}/\pt^{\JPsi}) \cdot L_{xy}$
and is required to be in the range from $-$0.05 to 0.1\cm.
The separation significance is defined as the ratio of the magnitude
of the three-dimensional vector $\Delta \vec{r}$ between the two reconstructed
\JPsi vertices
and the uncertainty of the distance measurement, $\sigma_{\Delta \vec{r}}$
(which includes the uncertainty in the vertex position, as determined by the
Kalman filter technique, and the uncertainty of the muon track fit):
$\delta d \equiv \abs{\Delta \vec{r}} / \sigma_{\Delta \vec{r}}$.  The requirement
$\delta d < 8$ is imposed.  From a data sample of pp collisions
corresponding to an integrated luminosity of
4.73\fbinv \cite{CMS-PAS-SMP-12-008}, 1043 candidate events
containing a \JPsi pair are found.

The kinematics of the $\JPsi\,\JPsi\to 4\mu$ final state is sensitive to
the underlying physics of production and decay, and this analysis probes
a higher-$\pt$ region of \JPsi pair production than previous experiments.
Therefore, the dependence on production model assumptions is minimized.
Given the relatively small number of events in the final-analysis
event sample it was affordable to calculate acceptance and efficiency
corrections on an event-by-event basis using the measured \JPsi and muon momenta. The procedure has the merit of not depending on
assumptions regarding correlations between production observables.

The muon acceptance is evaluated by generating a large number
of simulated decays starting from the reconstructed four momenta
of the two \JPsi mesons in an event.
The acceptance correction, $a_i$, for a given event $i$ is the
number of times all four muons survive the acceptance criteria,
listed in Eqs.~(\ref{eq:tpselect}) and (\ref{eq:looseselect}),
divided by the total number of trials for the event.
The angle of the decay muons with respect to the direction of flight
of the parent \JPsi, in the \JPsi rest frame, is assumed to be
isotropically distributed.  Deviations from this assumption are
considered and discussed later.
The event-by-event acceptance-correction procedure is
evaluated with both SPS and DPS MC simulation samples.
For each sample of $N$ events within the \JPsi phase space,
the muon acceptance criteria are applied to obtain a sample of
accepted events.  For each of the surviving events $i$,
the corresponding $a_i$ is obtained as described above. The corrected
number of signal events within the \JPsi phase space, $N^\prime$,
is then calculated as a sum over the survivors, $N^\prime = \sum_i 1/a_i$.
The difference between $N$ and $N^\prime$ is used to estimate
the systematic uncertainty in the method.

The efficiency correction is also determined on a per-event basis
by repeatedly generating \JPsi pair events where the
generated muon momenta are the measured muon momenta from the
reconstructed event. The event is then subjected to the complete
CMS detector simulation and reconstruction chain.
The efficiency correction, $\epsilon_i$, for a measured event $i$ is the
fraction of simulated events that pass the trigger and reconstruction
requirements.
The number of efficiency-corrected events is then given as
$\sum_i 1/\epsilon_i$, summed over the events that survive the
trigger and reconstruction requirements.  An average
efficiency for the sample in bins of the observables, $\Delta x$, is
obtained as the number of events that survive the trigger and reconstruction
requirements, divided by the number of efficiency-corrected events.
The method is evaluated with samples of reconstructed SPS
and DPS \JPsi pair MC simulation events.
For comparison, the average efficiency is alternatively determined from the SPS and DPS
MC simulation samples with simulated muon momenta. The average efficiency is
then given as the number of events surviving the trigger and reconstruction
criteria, divided by the number of events originally generated in the \JPsi phase space and muon acceptance region. In contrast to the first method,
this efficiency calculation is based on true muon momenta. The difference
between these two average efficiencies is due to the resolution of the
detector and is accounted for by a scaling factor which is in close agreement
between the two production models.

\section{Signal yield}
\label{sec:fit_technique}

An extended maximum likelihood method is performed to separate the 
signal from background contributions in the data sample. The signal 
weights $s_i$ in Eq.~(\ref{eq:cross-section}) are derived with the 
sPlot technique~\cite{sPlot}. The signal yield resulting from the 
fit is equal to the sum of the $s_i$. These weights are used to obtain 
the signal distribution in bins of kinematic variables that quantify 
the \JPsi pair production. Correlations between fit variables and 
production observables are found to be negligible from simulated samples.
Four kinematic variables are selected to discriminate the \JPsi pair signal
from the background: (i) the $\mu^+\mu^-$ invariant mass
of the higher-$\pt$ $\JPsi$, $M^{(1)}_{\mu\mu}$, (ii) the $\mu^+\mu^-$ invariant
mass of the lower-$\pt$ $\JPsi$, $M^{(2)}_{\mu\mu}$, (iii) the proper transverse decay
length of the higher-$\pt$ $\JPsi$, $ct_{xy}$, and (iv) the
separation significance, $\delta d$, between the two \JPsi candidates.
Five categories of events are identified: 
\begin{enumerate}
\item events containing a real prompt \JPsi pair (sig), 
\item background from at least one nonprompt
\JPsi meson, mostly from B-meson decays (nonprompt), 
\item the higher-$\pt$ prompt \JPsi and two unassociated muons 
that have an invariant mass within the \JPsi mass window,
\item the lower-$\pt$ prompt \JPsi and two unassociated muons 
that have an invariant mass within the \JPsi mass window, and
\item four unassociated muons (combinatorial-combinatorial).  
\end{enumerate}
The categories 3 and 4 have a common yield (\JPsi-combinatorial), and 
the parameter $f$ is defined as their relative fraction.
The likelihood function for event $j$ is obtained
by summing the product of the yields $n_{i}$ and the probability density
functions (PDFs) for the four kinematic variables $P_i(M^{(1)}_{\mu\mu})$,
$Q_i(M^{(2)}_{\mu\mu})$,
$R_i(ct_{xy})$, $S_i(\delta d)$ with the shape parameters for each of
the five event categories $i$.
The likelihood for each event $j$ is given as:
\begin{linenomath}
\begin{equation}
\begin{aligned}
{\ell}_j & = n_\text{sig}\left[ P_1 \cdot Q_1 \cdot R_1 \cdot S_1 \right]
 + n_\text{nonprompt}\left[ P_2\cdot Q_2\cdot R_2\cdot S_2 \right] \\
 & + n_{\JPsi\text{-combinatorial}}
[
 f \cdot P_3 \cdot Q_3 \cdot R_3 \cdot S_3 +  (1-f) \cdot P_4\cdot Q_4 \cdot R_4 \cdot S_4 ] \\
 & + n_\text{combinatorial-combinatorial}\left[ P_5 \cdot Q_5 \cdot R_5\cdot S_5 \right].
\end{aligned}
\end{equation}
\end{linenomath}
The yields $n_{i}$ are determined by minimizing the quantity
$-\ln{\mathcal{L}}$~\cite{roofit}, where $\mathcal{L} = \prod_j {\ell}_j$.

According to the signal MC simulation, the invariant mass and $ct_{xy}$ of the
higher-\pt \JPsi are correlated by about 13\%.  All other correlations between event
variables are below 5\%. Therefore, the parameterization for each variable is
independently determined. Several parameterizations for each distribution
are considered, and the simplest function with the least number of parameters
necessary to adequately describe the observed distribution is selected as
the PDF.  For parameterizations that result in equally good descriptions of
the data (as measured by the $\chi^2$ of the fit of the distribution in data
for a given variable), the difference in signal yields is used as a measure
of the systematic uncertainty.

For the likelihood fit, the sum of two Gaussian functions with a common mean is
used to parameterize the signal \JPsi invariant mass PDFs $P_1$ and $Q_1$; the
same parameters are used to describe the nonprompt components $P_2$ and $Q_2$,
and the \JPsi part of the \JPsi-combinatorial cases $P_3$ and $Q_4$.  The
widths of the Gaussian functions are fixed to the best-fit values obtained
in simulation samples.  A sum of two Gaussians is also used
to describe the signal $ct_{xy}$ PDF $R_1$. The nonprompt background
distribution $R_2$ is fit by an exponential function convolved with a single
Gaussian. The separation significance PDFs for the signal and nonprompt
components, $S_1$ and $S_2$, are parameterized with a single Gaussian convolved with
an exponential function.  Simulated event samples are used to parameterize
the prompt and nonprompt $ct_{xy}$ and $\delta d$ distributions.
The distributions of the signal variables as predicted by the simulation of
SPS production agree with those from DPS production.

Combinatorial background shapes are obtained directly from data.
Two $M_{\mu\mu}$ sideband regions are defined in the ranges
$[2.85,3]$ and $[3.2,3.35]\GeVcc$, adjacent to the signal region
defined as $[3,3.2]\GeVcc$, and PDF parameters are estimated
from fits to combinations of samples in data where only one or neither of
the \JPsi candidates originate from the signal region. The mass distributions
are parameterized under the assumption that they only contain contributions
from true \JPsi candidates and combinatorial background. Third-order Chebyshev
polynomial functions are used to describe the combinatorial components of each
invariant mass PDF $Q_3$ and $P_4$ in the partially combinatorial and completely
combinatorial category. In the latter case, it is required that $P_5$ equals
$P_4$ and $Q_5$ equals $Q_3$. A sum of two Gaussians is used
for $R_{3-5}$, and a Landau function plus a first-order
Chebyshev polynomial is used to parameterize $S_{3-5}$.

\begin{figure*}[!ht]
\centering
    \includegraphics[width=0.49\textwidth]{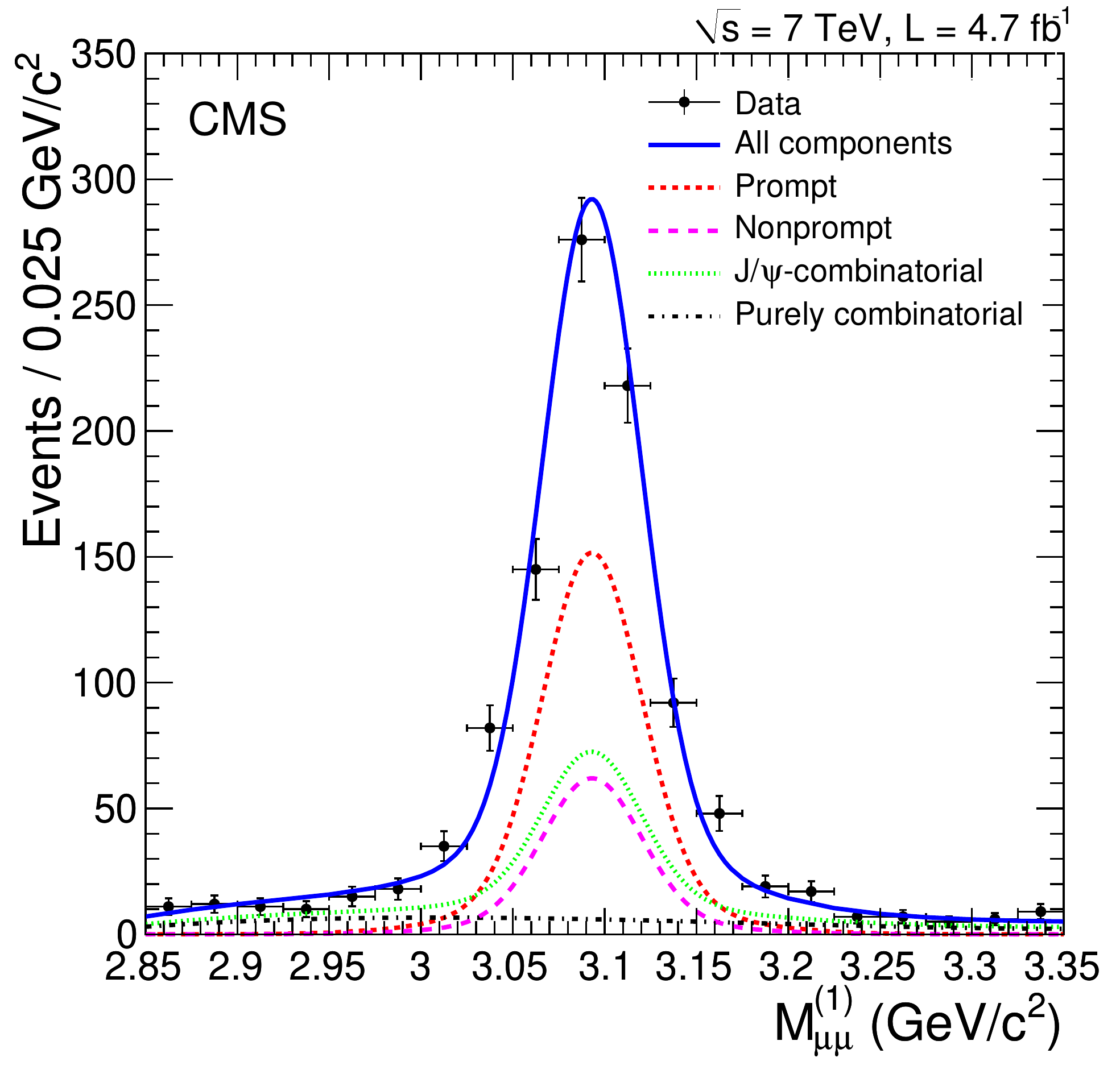}
    \label{fig:fit1}
    \includegraphics[width=0.49\textwidth]{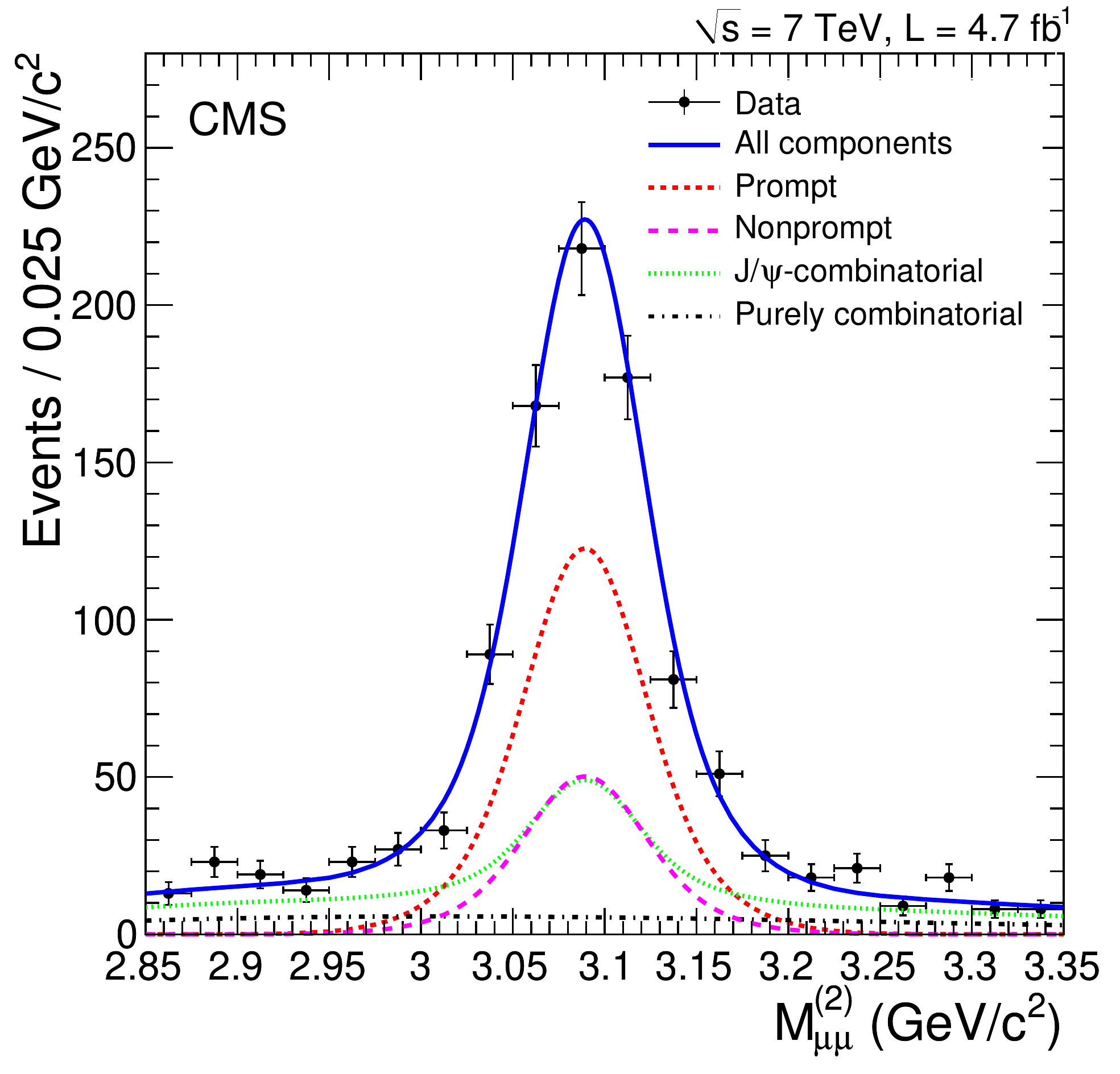}
    \label{fig:fit2}
    \includegraphics[width=0.49\textwidth]{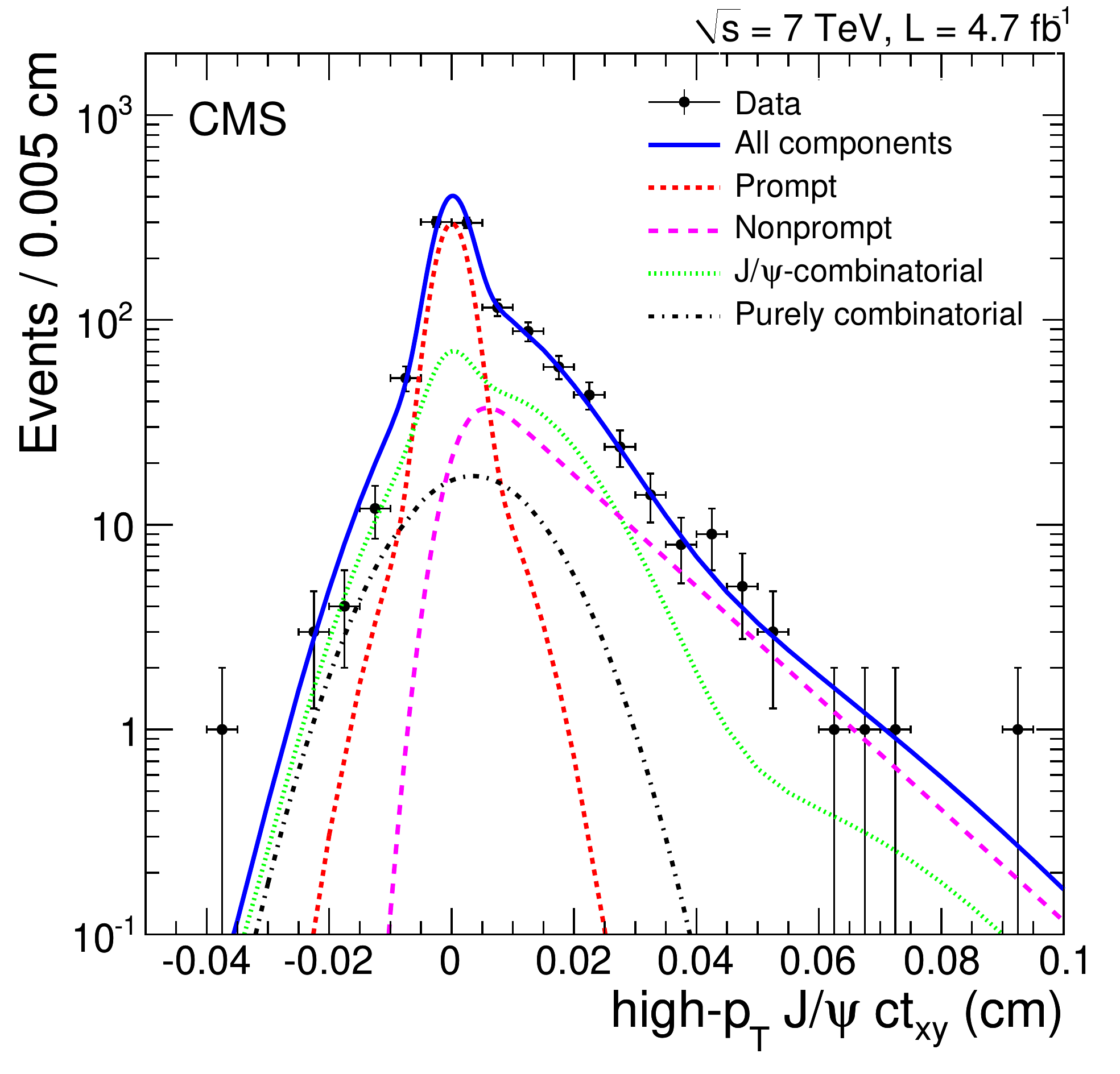}
    \label{fig:fit3}
    \includegraphics[width=0.49\textwidth]{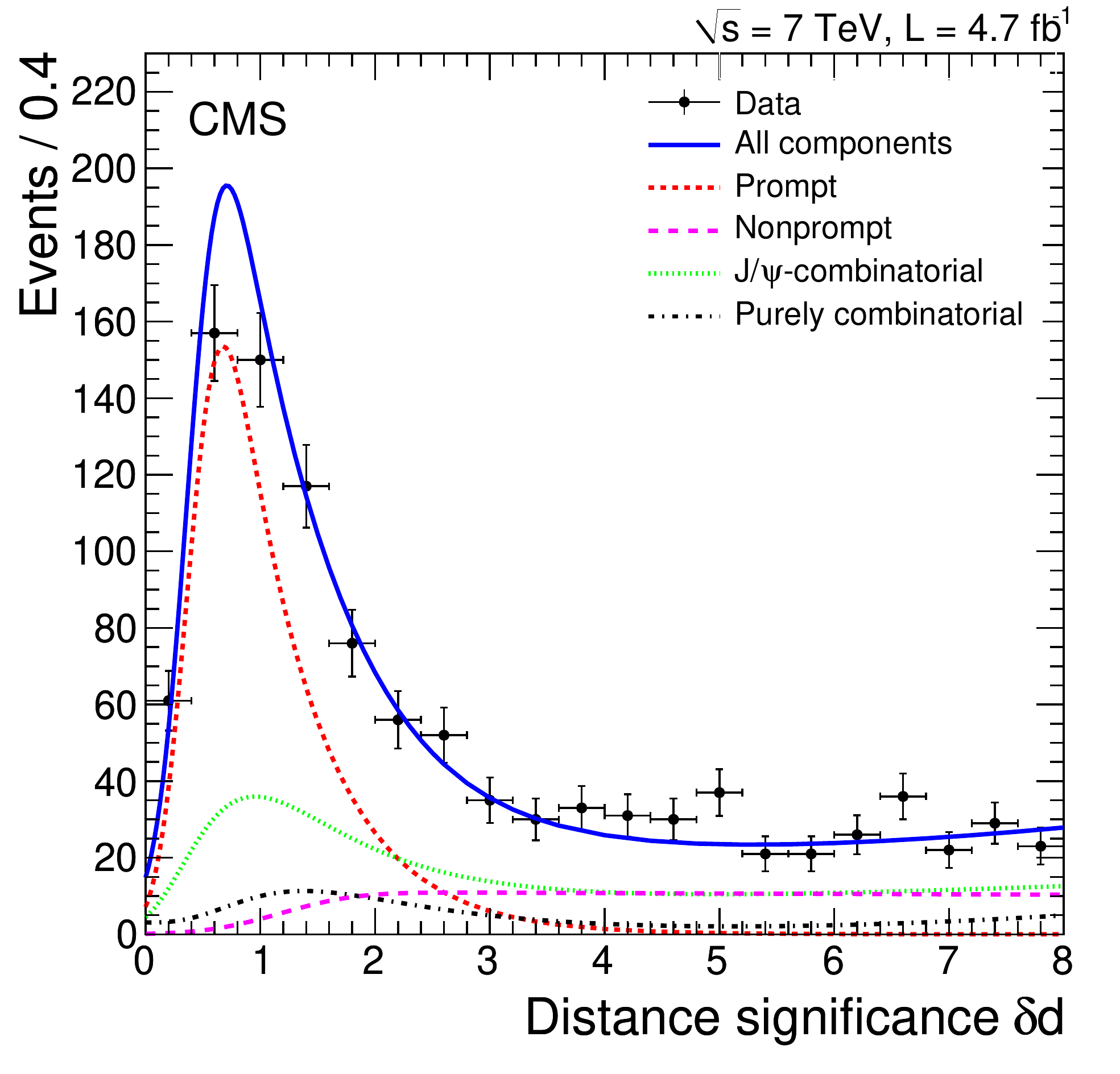}
    \label{fig:fit4}
    \caption{Distributions of $M_{\mu\mu}^{(1)}$~(top left),
$M_{\mu\mu}^{(2)}$~(top right), $ct_{xy}$~(bottom left), and distance
significance $\delta d$ (bottom right) for the candidate events and the
projections of the fit results.
The data are shown as points with the vertical error bars representing the
statistical uncertainty.  The fit result to the full sample is shown as a
solid line.  Individual contributions from
the various categories are shown in different line styles: signal (short dashes),
nonprompt background (long dashes), $\JPsi$-combinatorial components
(dots), and the pure combinatorial component (dashes and dots).
}
    \label{fig:fit}
\end{figure*}

The final fit is performed on the full data sample.  The mean values of the
central Gaussian functions of the two $\mu^+\mu^-$ invariant-mass distributions
are left free, as is the proper decay time of the nonprompt component.
The fit yields $n_\text{sig} = 446 \pm 23$ signal events.
Figure~\ref{fig:fit} shows the distributions of the event variables from
data with the fit result superimposed.  The fit is validated by repeatedly
generating simulated samples from the PDFs for all
components and no bias is found. Furthermore, the robustness of the fit is
probed by adding combinations of simulated signal and background events
to the data set. To ensure that the cross section determination is insensitive
to changing conditions, the distributions of the variables used in the
likelihood fit are compared in subsets of events.  Event variable
distributions from events containing six reconstructed primary vertices or
fewer agree with distributions in events containing more than six primary
vertices (within statistical uncertainties).  The behavior is confirmed with
MC simulation signal samples generated with and without pileup contributions.
The variable distributions also agree between the two major 2011 data-taking periods.

\section{Systematic uncertainties}
\label{sec:syst}

The uncertainty in the \JPsi dimuon {branching fraction} is taken from
the world average~\cite{PDG2012} (2\% when added
linearly).  The systematic uncertainty corresponding to the integrated
{luminosity} normalization is estimated in previous studies (2.2\%)~\cite{CMS-PAS-SMP-12-008}.  Simulated event samples based on SPS and DPS production models are
used to estimate the uncertainty in the event-by-event {acceptance
correction} method: $N$ simulated events are subjected to the acceptance
criteria, and the event-based acceptance correction is applied to arrive
at a corrected yield, $N^\prime$.  The uncertainty is taken as half of the
relative difference between the two yields, $N$ and $N^\prime$. The larger
value among the SPS- and DPS-based samples is quoted (1.1\%).  The precision
of the event-based {efficiency correction} is limited by the number of
reconstructed events, $n_{\text{reco},i}$, found after the substitution
process for each event $i$.  The cross section is recalculated by
repeatedly varying $n_{\text{reco},i}$ according to Gaussian functions with
standard deviation $\sqrt{\smash[b]{n_{\text{reco},i}}}$.  The standard deviation of the
resulting cross section distribution is used as an estimate of the
uncertainty in the efficiency calculation (4.4\%).  The relative
{efficiency scaling factor} is determined from SPS and DPS MC
simulation samples, representing very different
scenarios of \JPsi pair kinematics.  The uncertainty due to model
dependence of the scaling factor is defined as the difference in the
cross section between either model and the average of the two (0.2\%).
The small uncertainty demonstrates that there is little
overall model dependence.

The {muon track reconstruction} efficiency is derived from simulated
events.  The uncertainty is estimated from data and simulation samples
that contain at least one reconstructed \JPsi. For each muon in an event, the
tracking efficiency in data and simulation is obtained as a function
of the measured muon pseudorapidity~\cite{CMS-PAS-TRK-10-002}.
The relative uncertainty is defined as the absolute difference between the
data- and simulation-based values divided by the data-based value. Individual
muon uncertainties are added linearly per event (3.0\%) since correlations
between the muons are not taken into account.

The efficiency to trigger and reconstruct \JPsi pair
events relies on {detector simulation}. To evaluate the uncertainty
event-based efficiency values are instead constructed from single-muon
efficiencies.  The single-muon efficiencies are determined by applying a
``tag-and-probe'' method~\cite{MUO-10-004} to control samples in data and
simulation that contain single \JPsi decays to muons.  Hence, correlations among the two \JPsi mesons in the event are neglected.
The difference in the signal yield in data when corrected with efficiencies found
from either data or simulation is used to measure the uncertainty.
The event-based efficiency correction is defined as the
product of the event's trigger efficiency, given that all muons are found
offline, and the event efficiency for reconstructing, identifying, and
selecting offline all four muons in an event. The trigger efficiency is
calculated from the single-muon trigger efficiencies and the dimuon
vertexing efficiency as the trigger requires at least three reconstruced
muons, two of which must be fit to a \JPsi vertex.
The offline reconstruction efficiency for a single muon is given as the
product of the tracking efficiency, muon identification efficiency, and the
efficiency to pass the offline quality criteria.  All muon efficiencies are
obtained as a function of muon $\pt$\ and $\eta$\ from previous
studies~\cite{MUO-10-004}.  The probability to successfully fit both
vertices in an event is greater than 99.9\% for SPS and 99.6\% for
DPS simulation samples.  Therefore, the offline event reconstruction
efficiency is considered to be entirely a product of the muon reconstruction
efficiencies.  The largest deviation of the corrected signal yield using
the single-muon efficiency values from data control samples compared
to simulation is chosen as a conservative measure of the uncertainty
(6.5\%).

All {PDF parameters} that are fixed for the maximum likelihood
fit are varied by their uncertainty, as determined from the fits to the data
sidebands and MC simulation samples.  The prompt $ct_{xy}$ distribution
is also parameterized using a sum of three Gaussians.
Alternative fit shapes such as third-order polynomials or exponential
functions are used for the background models.
A Crystal Ball function~\cite{ref:crystalball} is considered as an
alternative to
the parameterization of the \JPsi invariant-mass distribution.  A resolution
function convolved with an exponential function is considered for
the separation significance of the combinatorial background components. The
largest difference in signal yields between fits with different shape
parameterizations is taken as the uncertainty from the PDFs (0.6\%).
To evaluate the dependence of the PDF
parameterization on the {production model}, both reconstructed DPS and SPS
samples are used.  The difference in signal yields between fits with those two
PDF sets is considered as an uncertainty (0.1\%).
The total systematic uncertainty is calculated as the sum in quadrature of
the individual uncertainties (9.0\%).  The individual relative uncertainties
for the total cross section are listed in Table~\ref{tab:systematics}.
The systematic uncertainty for each differential cross section is also
evaluated on a per-bin basis for all uncertainties due to the acceptance and
efficiency corrections.

\begin{table}[htbp]
\centering
\topcaption{Summary of relative systematic uncertainties in the
\JPsi pair total cross section.}
\begin{tabular}{lc} \hline
Source                         & Relative uncertainty [\%] \\
\hline
Branching fraction             &  2.0 \\
Integrated luminosity          &  2.2 \\
Acceptance correction          &  1.1 \\
Efficiency correction          &  4.4 \\
Efficiency scaling factor      &  0.2 \\
Muon track reconstruction      &  3.0 \\
Detector simulation            &  6.5 \\
PDF parameters                 &  0.6 \\
Production model               &  0.1 \\
\hline
Total                          & 9.0 \\
\hline
\end{tabular}
\label{tab:systematics}
\end{table}

To study the effect of nonisotropic \JPsi decay into muons on the
measured cross section, the event-based acceptance is determined using
extreme scenarios.
Defining $\theta$ as the angle between the $\mu^+$ direction in the
\JPsi rest frame and the \JPsi direction in the pp center-of-mass
frame, the angular distribution of decay muons is parameterized as:
$f(\theta) = 1 + \lambda \cos^2\theta$,
where $\lambda$ is a polarization observable~\cite{Chao:2012iv},
with $\lambda = 0$ corresponding to an isotropic
\JPsi decay.  Compared to the $\lambda = 0$ case, the total cross section
is 31\% lower for $\lambda = -1$ and 27\%  higher for $\lambda = +1$.
The differential cross section measurements for $\lambda = \pm 1$ lie
within the statistical uncertainties of the $\lambda = 0$ case when scaled
to the same total cross section, indicating that different polarization
assumptions do not affect the shapes of the cross section distributions.
Once the value of $\lambda$ has been measured, it can be
used in the acceptance calculation to mitigate
this source of uncertainty.

\section{Results}
\label{sec:final}
\begin{figure}[htbp]
  \begin{center}
    \includegraphics[width=0.49\textwidth]{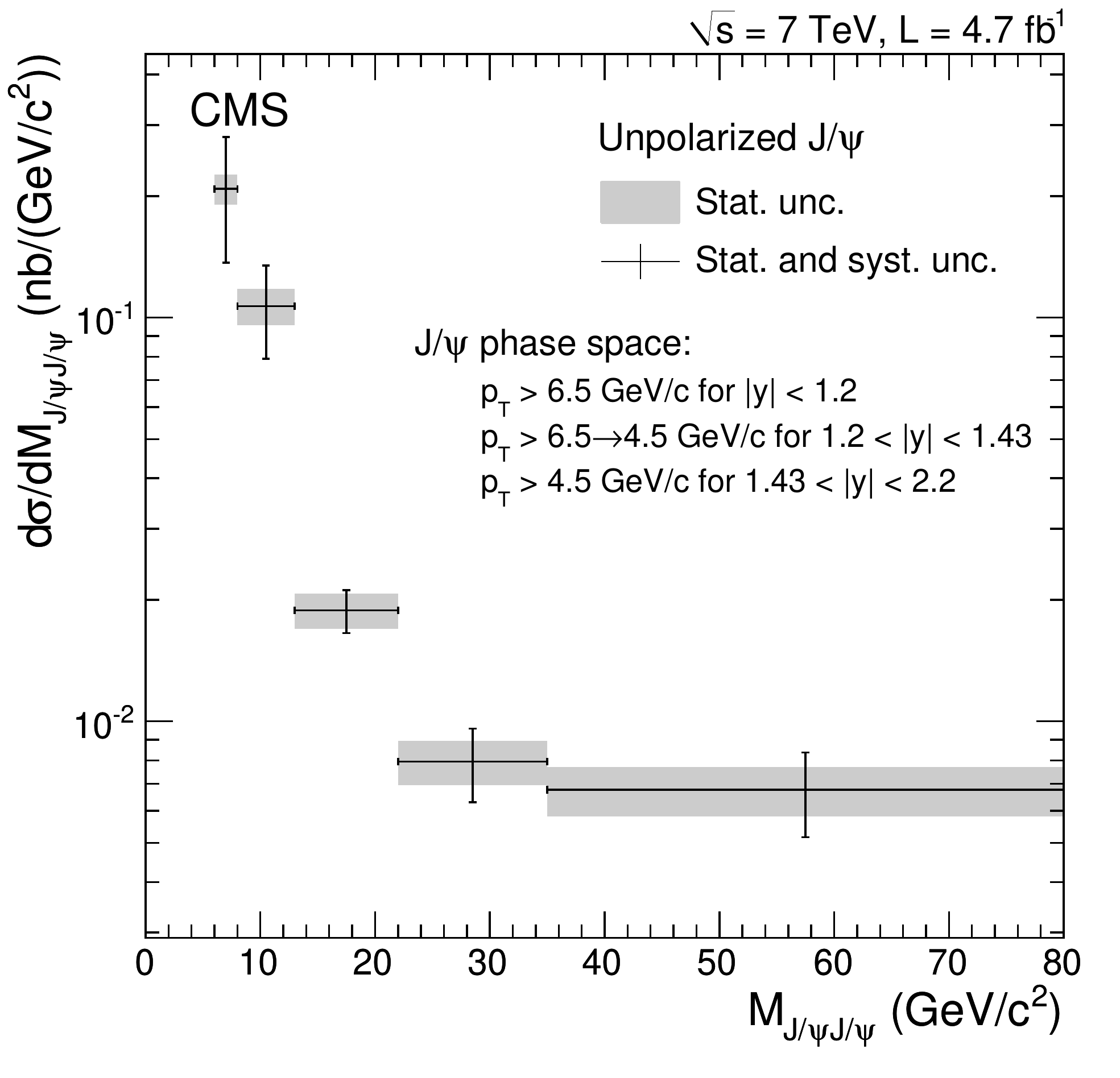}
    \includegraphics[width=0.49\textwidth]{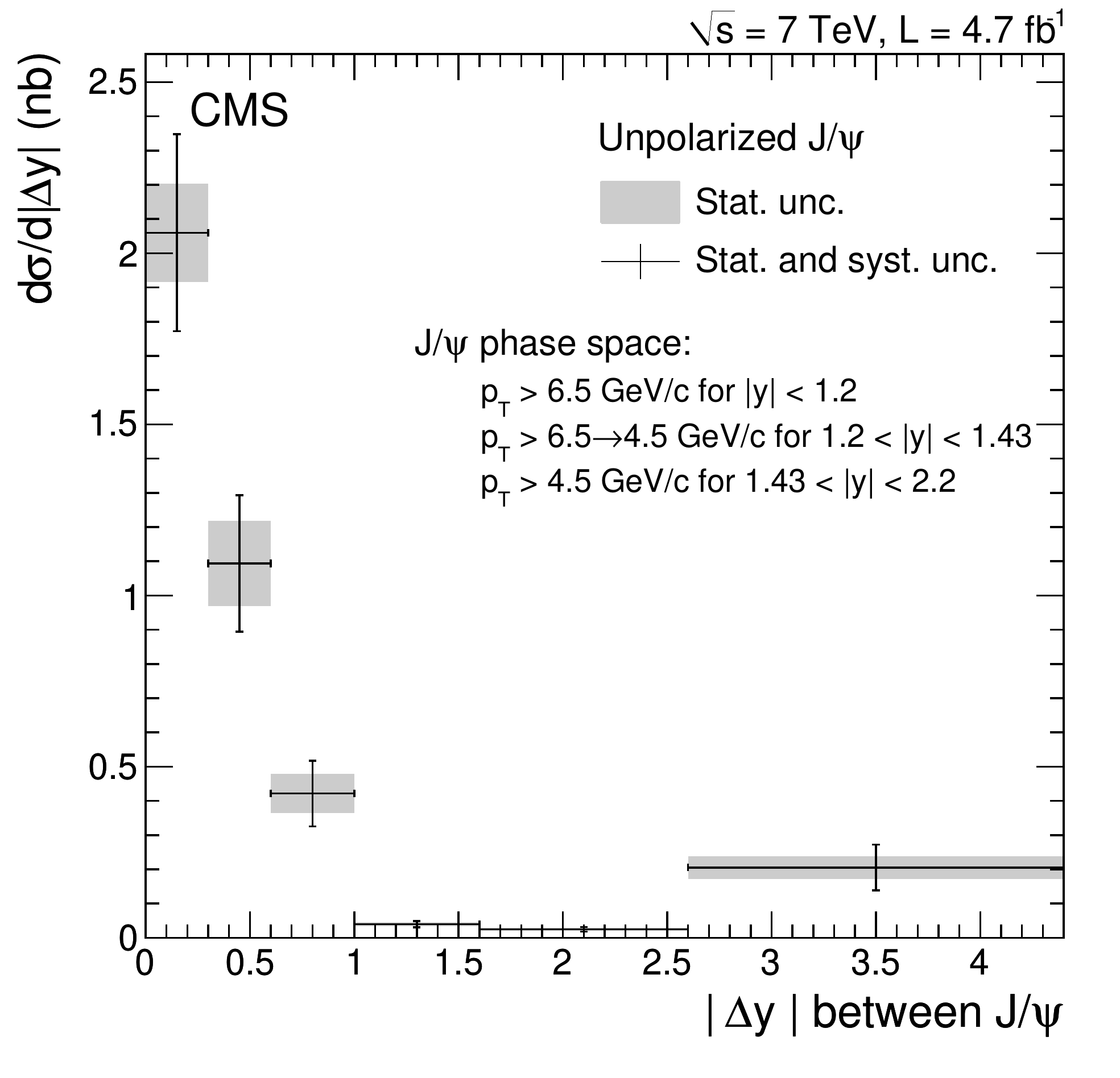}
    \includegraphics[width=0.49\textwidth]{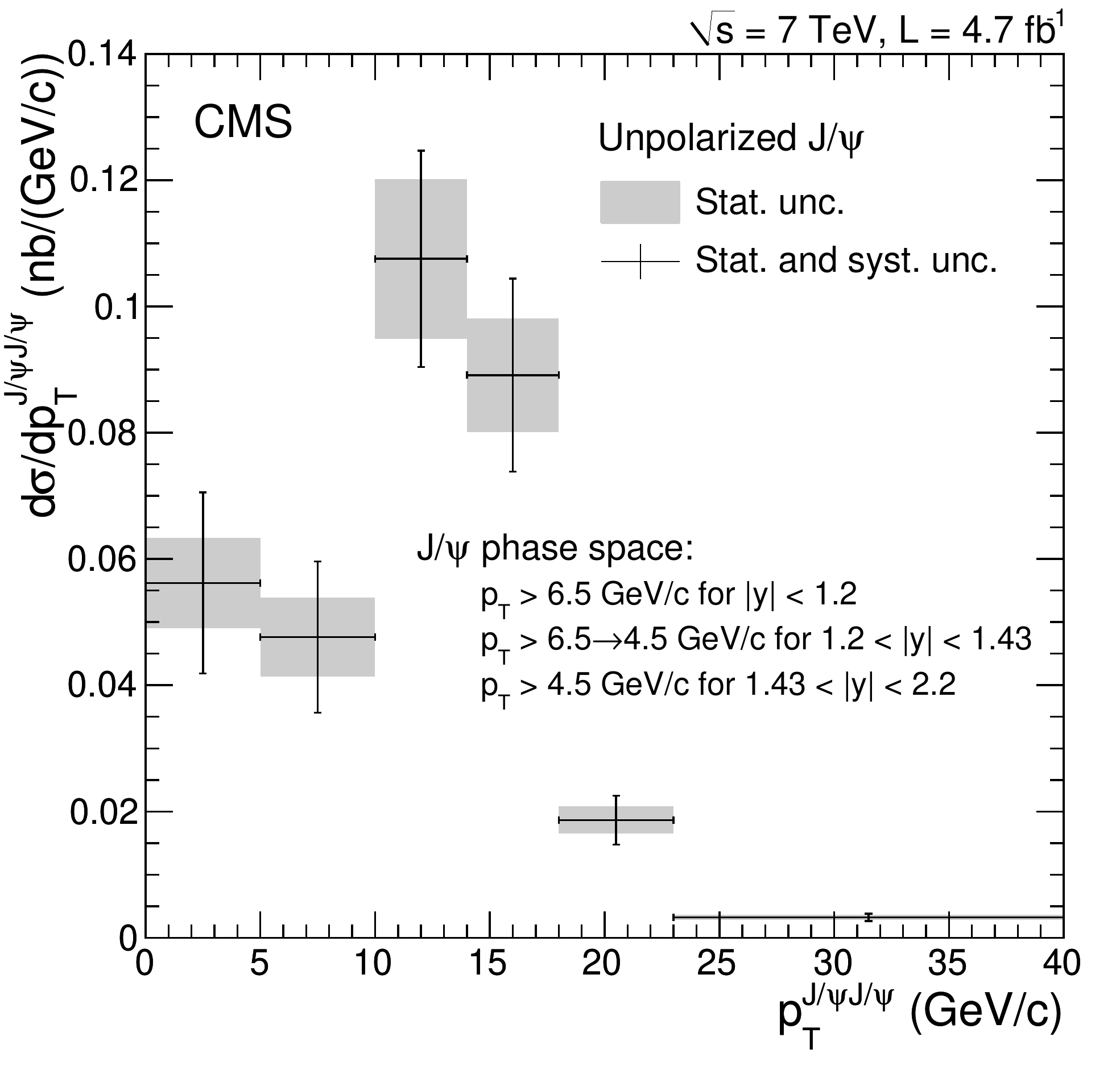}
    \caption{Differential cross section for prompt \JPsi pair production
as a function of the \JPsi pair invariant mass ($M_{\JPsi\,\JPsi}$, top left),
the absolute rapidity difference between \JPsi mesons ($\abs{\Delta y}$, top
right), and the \JPsi pair transverse momentum ($\pt^{\JPsi\,\JPsi}$, bottom),
over the \JPsi phase space given in the figure, assuming unpolarized \JPsi production. The shaded regions represent the statistical uncertainties only,
and the error bars represent the statistical and systematic uncertainties
added in quadrature.
}
    \label{fig:dsigdx}
  \end{center}
\end{figure}

The total cross section obtained by summing over the sample on an
event-by-event basis and assuming unpolarized prompt
\JPsi pair production is
\begin{linenomath}
\begin{equation}
\sigma({\Pp\Pp}\to \JPsi\, \JPsi+X) = 1.49 \pm 0.07 \pm 0.13\unit{nb},
\end{equation}
\end{linenomath}
with statistical and systematic uncertainties shown, respectively.
For the measurement, the values $\mathcal{L} = 4.73 \pm 0.10$\fbinv~\cite{CMS-PAS-SMP-12-008}
and $BF(\JPsi \to \Pgmp\Pgmm)$
= (5.93 $\pm$ 0.06)\% \cite{PDG2012} are used.  The differential cross
section as a function of the \JPsi pair invariant mass ($M_{\JPsi\,\JPsi}$),
the absolute rapidity difference between \JPsi mesons ($\abs{\Delta y}$), and
the \JPsi pair transverse momentum ($\pt^{\JPsi\,\JPsi}$) is shown in
Fig.~\ref{fig:dsigdx}.
The observed differential cross section is not only a result of the kinematics
of \JPsi pair production, but also of the \JPsi phase-space window (given in
the figures) available for measurement.
The corresponding numerical values are summarized
in Tables~\ref{tab:dsigdmTable}, \ref{tab:dsigdyTable},
and~\ref{tab:dsigdptTable}, respectively.

\begin{table}[h!]
\centering
\topcaption{Differential cross section in bins of the \JPsi pair invariant
mass ($M_{\JPsi\,\JPsi}$).  The uncertainties shown are statistical first,
then systematic.  }
\begin{tabular*}{0.49\textwidth}{@{\extracolsep{\fill}}ll} \hline
$M_{\JPsi\,\JPsi}$  (\GeVccns)& $\rd\sigma / \rd{}M_{\JPsi\,\JPsi}$  (nb/(\GeVccns)) \\ \hline
6--8 & $0.208 \pm 0.018 \pm 0.069$ \\
8--13 & $0.107 \pm 0.011 \pm 0.025$ \\
13--22 & $0.019 \pm 0.002 \pm 0.001$ \\
22--35 & $0.008 \pm 0.001 \pm 0.001$ \\
35--80 & $0.007 \pm 0.001 \pm 0.001$ \\
\hline
\end{tabular*}
\label{tab:dsigdmTable}
\end{table}

\begin{table}[h!]
\centering
\topcaption{Differential cross section in bins of the
absolute rapidity difference between \JPsi mesons ($\abs{\Delta y}$).
The uncertainties shown are statistical first, then systematic.}
\begin{tabular*}{0.49\textwidth}{@{\extracolsep{\fill}}ll} \hline
$\abs{\Delta y}$ & $\rd\sigma / \rd\abs{\Delta y}$  (nb) \\ \hline
0--0.3 & $2.06 \pm 0.14 \pm 0.25$ \\
0.3--0.6 & $1.09 \pm 0.13 \pm 0.16$ \\
0.6--1 & $0.421 \pm 0.057 \pm 0.077$ \\
1--1.6 & $0.040 \pm 0.006 \pm 0.006$ \\
1.6--2.6 & $0.025 \pm 0.005 \pm 0.005$ \\
2.6--4.4 & $0.205 \pm 0.033 \pm 0.058$ \\
\hline
\end{tabular*}
\label{tab:dsigdyTable}
\end{table}

\begin{table}[h!]
\centering
\topcaption{Differential cross section in bins of the transverse momentum
of the \JPsi pair ($\pt^{\JPsi\,\JPsi}$).  The uncertainties shown are
statistical first, then systematic.}
\begin{tabular*}{0.49\textwidth}{@{\extracolsep{\fill}}ll} \hline
$\pt^{\JPsi\,\JPsi}$ ($\GeVc$) & $\rd\sigma / \rd\pt^{\JPsi\,\JPsi}$~(nb/(\GeVcns)) \\ \hline
0--5 & $0.056 \pm 0.007 \pm 0.012$ \\
5--10 & $0.048 \pm 0.006 \pm 0.010$ \\
10--14 & $0.108 \pm 0.013 \pm 0.012$ \\
14--18 & $0.089 \pm 0.009 \pm 0.012$ \\
18--23 & $0.019 \pm 0.002 \pm 0.003$ \\
23--40 & $0.003 \pm 0.001 \pm 0.001$ \\
\hline
\end{tabular*}
\label{tab:dsigdptTable}
\end{table}

A search for the $\eta_b$ is performed by examining
the \JPsi pair invariant-mass distribution around the
nominal $\eta_b$ mass~\cite{PDG2012}, before efficiency and acceptance
corrections.  From samples of simulated \JPsi pair events produced
via SPS or DPS, the acceptance times efficiency is found to be nearly
linear in the mass interval 8.68--10.12\GeVcc.  The reconstructed
Gaussian width of the $\eta_b$ is 0.08\GeVcc, as determined from a
\JPsi pair MC simulation sample generated according to a Breit--Wigner
function with the nominal $\eta_b$ mass and width~\cite{PDG2012}.
The signal search interval
9.16--9.64\GeVcc corresponds to three standard deviations on each
side of the mean mass value.  Two sideband regions of the same width
as the signal region are defined as the intervals 8.68--9.16\GeVcc and 9.64--10.12\GeVcc.
A first-degree polynomial is used to fit the number of events in the
sideband regions. Extrapolating these yields to the signal region predicts
$15\pm 4$ nonresonant events.  The total number of \JPsi pair events in this
region in data is 15. Hence, no significant $\eta_b$ contribution is observed.

\section{Summary}
\label{sec:interpretation}

A signal yield of $446\pm 23$ events for the production of prompt
\JPsi meson pairs has been observed with
the CMS detector in pp collisions at $\sqrt{s}=7$\TeV
from a sample corresponding to an integrated luminosity of
$4.73 \pm 0.10$\fbinv.  A data-based method has been used to correct for
the acceptance and efficiency, minimizing the model dependence of the cross
section determination.  The total cross section of prompt \JPsi pair
production measured within a phase-space region defined by the
individual \JPsi \pt and rapidity is found to be
$1.49 \pm 0.07\stat \pm 0.13\syst$\unit{nb},
where unpolarized production is assumed.
Differential cross sections have been obtained in bins of the
\JPsi pair invariant mass, the absolute rapidity difference between the two
\JPsi mesons, and the \JPsi pair transverse momentum.
These measurements probe \JPsi pair production at higher \JPsi \pt and more central rapidity than the LHCb measurement~\cite{Aaij:2012dz},
providing for the first time information about a kinematic region
where color-octet \JPsi states and higher-order corrections play a
greater role in production. The differential cross section in
bins of $\abs{\Delta y}$ is sensitive to DPS contributions to
prompt \JPsi pair production. The differential cross section
decreases rapidly as a function of $\abs{\Delta y}$. However, a non-zero
value is measured in the $\abs{\Delta y}$ bin between 2.6 and 4.4.
Current models predict that this region can be populated via DPS
production~\cite{Baranov:2011ch,Baranov:2013,Kom:2011bd}.

There is no evidence for the $\eta_b$ resonance in the \JPsi pair invariant-mass distribution above the background expectations derived from the $\eta_b$ sideband regions.
Since models describing the
nonresonant \JPsi pair production in the CMS \JPsi phase-space
window are not available, an upper limit on the production cross section times branching fraction for $\eta_b\to \JPsi\, \JPsi$ cannot be obtained.

Model descriptions of \JPsi pair production at higher
\pt are crucial input to quantify nonresonant contributions in the search for new states at different center-of-mass energies. The cross section measurements presented here provide significant new information
for developing improved theoretical production models.

\section*{Acknowledgements}

We congratulate our colleagues in the CERN accelerator departments for the excellent performance of the LHC and thank the technical and administrative staffs at CERN and at other CMS institutes for their contributions to the success of the CMS effort. In addition, we gratefully acknowledge the computing centres and personnel of the Worldwide LHC Computing Grid for delivering so effectively the computing infrastructure essential to our analyses. Finally, we acknowledge the enduring support for the construction and operation of the LHC and the CMS detector provided by the following funding agencies: BMWFW and FWF (Austria); FNRS and FWO (Belgium); CNPq, CAPES, FAPERJ, and FAPESP (Brazil); MES (Bulgaria); CERN; CAS, MoST, and NSFC (China); COLCIENCIAS (Colombia); MSES and CSF (Croatia); RPF (Cyprus); MoER, ERC IUT and ERDF (Estonia); Academy of Finland, MEC, and HIP (Finland); CEA and CNRS/IN2P3 (France); BMBF, DFG, and HGF (Germany); GSRT (Greece); OTKA and NIH (Hungary); DAE and DST (India); IPM (Iran); SFI (Ireland); INFN (Italy); NRF and WCU (Republic of Korea); LAS (Lithuania); MOE and UM (Malaysia); CINVESTAV, CONACYT, SEP, and UASLP-FAI (Mexico); MBIE (New Zealand); PAEC (Pakistan); MSHE and NSC (Poland); FCT (Portugal); JINR (Dubna); MON, RosAtom, RAS and RFBR (Russia); MESTD (Serbia); SEIDI and CPAN (Spain); Swiss Funding Agencies (Switzerland); MST (Taipei); ThEPCenter, IPST, STAR and NSTDA (Thailand); TUBITAK and TAEK (Turkey); NASU and SFFR (Ukraine); STFC (United Kingdom); DOE and NSF (USA).

Individuals have received support from the Marie-Curie programme and the European Research Council and EPLANET (European Union); the Leventis Foundation; the A. P. Sloan Foundation; the Alexander von Humboldt Foundation; the Belgian Federal Science Policy Office; the Fonds pour la Formation \`a la Recherche dans l'Industrie et dans l'Agriculture (FRIA-Belgium); the Agentschap voor Innovatie door Wetenschap en Technologie (IWT-Belgium); the Ministry of Education, Youth and Sports (MEYS) of the Czech Republic; the Council of Science and Industrial Research, India; the HOMING PLUS programme of Foundation for Polish Science, cofinanced from European Union, Regional Development Fund; the Compagnia di San Paolo (Torino); and the Thalis and Aristeia programmes cofinanced by EU-ESF and the Greek NSRF.

\bibliography{auto_generated}

\providecommand{\href}[2]{#2}\begingroup\raggedright\begin{thebibliography}{10}%
\makeatletter
\providecommand{\hrefCMSnoop }[0]{\@secondoftwo}%
\makeatother
\providecommand{\doi}{\texttt{doi:}\begingroup \urlstyle{tt}\Url}

\bibitem{Kom:2011bd}
\hrefCMSnoop {} {C.~H. Kom, A.~Kulesza, and W.~J. Stirling, ``{Pair Production
  of \JPsi as a Probe of Double Parton Scattering at LHCb}'',} \textit{ Phys.
  Rev. Lett.} \textbf{ 107} (2011) 082002,
  \href{http://dx.doi.org/10.1103/PhysRevLett.107.082002}{\doi{10.1103/PhysRevLett.107.082002}},
\href{http://www.arXiv.org/abs/1105.4186}{\texttt{ arXiv:1105.4186}}.

\bibitem{Ko:2010xy}
\hrefCMSnoop {} {P.~Ko, C.~Yu, and J.~Lee, ``{Inclusive double-quarkonium
  production at the Large Hadron Collider}'',} \textit{ JHEP} \textbf{ 01}
  (2011) 070,
  \href{http://dx.doi.org/10.1007/JHEP01(2011)070}{\doi{10.1007/JHEP01(2011)070}},
\href{http://www.arXiv.org/abs/1007.3095}{\texttt{ arXiv:1007.3095}}.

\bibitem{Berger:2009cm}
\hrefCMSnoop {} {E.~L. Berger, C.~B. Jackson, and G.~Shaughnessy,
  ``{Characteristics and Estimates of Double Parton Scattering at the Large
  Hadron Collider}'',} \textit{ Phys. Rev.} \textbf{ D 81} (2010) 014014,
  \href{http://dx.doi.org/10.1103/PhysRevD.81.014014}{\doi{10.1103/PhysRevD.81.014014}},
\href{http://www.arXiv.org/abs/0911.5348}{\texttt{ arXiv:0911.5348}}.

\bibitem{Baranov:2011ch}
\hrefCMSnoop {} {S.~P. Baranov, A.~M. Snigirev, and N.~P. Zotov, ``{Double
  heavy meson production through double parton scattering in hadronic
  collisions}'',} \textit{ Phys. Lett.} \textbf{ B 705} (2011) 116,
  \href{http://dx.doi.org/10.1016/j.physletb.2011.09.106}{\doi{10.1016/j.physletb.2011.09.106}},
\href{http://www.arXiv.org/abs/1105.6276}{\texttt{ arXiv:1105.6276}}.

\bibitem{Baranov:2013}
S.~P. Baranov\hrefCMSnoop {} { {et~al.}, ``Interparticle correlations in the
  production of \JPsi\ pairs in proton-proton collisions'',} \textit{ Phys.
  Rev.} \textbf{ D 87} (2013) 034035,
  \href{http://dx.doi.org/10.1103/PhysRevD.87.034035}{\doi{10.1103/PhysRevD.87.034035}},
  \href{http://www.arXiv.org/abs/1210.1806}{\texttt{ arXiv:1210.1806}}.

\bibitem{Badier:1982ae}
\hrefCMSnoop {} {{NA3 Collaboration}, J.~Badier {et~al.}, ``{Evidence for $\psi
  \psi$ Production in $\pi^-$ Interactions at 150~{GeV}/$c$ and
  280~{GeV}/$c$}'',} \textit{ Phys. Lett.} \textbf{ B 114} (1982) 457,
\href{http://dx.doi.org/10.1016/0370-2693(82)90091-0}{\doi{10.1016/0370-2693(82)90091-0}}.

\bibitem{Badier:1985ri}
\hrefCMSnoop {} {{NA3 Collaboration}, J.~Badier {et~al.}, ``{$\psi \psi$
  Production and Limits on Beauty Meson Production from 400~{GeV}/$c$
  Protons}'',} \textit{ Phys. Lett.} \textbf{ B 158} (1985) 85,
\href{http://dx.doi.org/10.1016/0370-2693(85)90745-2}{\doi{10.1016/0370-2693(85)90745-2}}.

\bibitem{Humpert:1983yj}
\hrefCMSnoop {} {B.~Humpert and P.~Mery, ``{$\psi \psi$ production at collider
  energies}'',} \textit{ Z. Phys.} \textbf{ C 20} (1983) 83,
\href{http://dx.doi.org/10.1007/BF01577721}{\doi{10.1007/BF01577721}}.

\bibitem{Berezhnoy:2011xy}
\hrefCMSnoop {} {A.~V. Berezhnoy, A.~K. Likhoded, A.~V. Luchinsky, and A.~A.
  Novoselov, ``{Double \JPsi -meson Production at LHC and 4$c$-tetraquark
  state}'',} \textit{ Phys. Rev.} \textbf{ D 84} (2011) 094023,
  \href{http://dx.doi.org/10.1103/PhysRevD.84.094023}{\doi{10.1103/PhysRevD.84.094023}},
\href{http://www.arXiv.org/abs/1101.5881}{\texttt{ arXiv:1101.5881}}.

\bibitem{Qiao:2009kg}
\hrefCMSnoop {} {C.-F. Qiao, L.-P. Sun, and P.~Sun, ``{Testing Charmonium
  Production Mechamism via Polarized \JPsi\ Pair Production at the LHC}'',}
  \textit{ J. Phys.} \textbf{ G 37} (2010) 075019,
  \href{http://dx.doi.org/10.1088/0954-3899/37/7/075019}{\doi{10.1088/0954-3899/37/7/075019}},
\href{http://www.arXiv.org/abs/0903.0954}{\texttt{ arXiv:0903.0954}}.

\bibitem{Campbell:2007ws}
\hrefCMSnoop {} {J.~M. Campbell, F.~Maltoni, and F.~Tramontano, ``{QCD
  corrections to \JPsi\ and $\Upsilon$ production at hadron colliders}'',}
  \textit{ Phys. Rev. Lett.} \textbf{ 98} (2007) 252002,
  \href{http://dx.doi.org/10.1103/PhysRevLett.98.252002}{\doi{10.1103/PhysRevLett.98.252002}},
\href{http://www.arXiv.org/abs/hep-ph/0703113}{\texttt{ arXiv:hep-ph/0703113}}.

\bibitem{Artoisenet:2007xi}
\hrefCMSnoop {} {P.~Artoisenet, J.~P. Lansberg, and F.~Maltoni,
  ``{Hadroproduction of \JPsi\ and $\Upsilon$ in association with a heavy-quark
  pair}'',} \textit{ Phys. Lett.} \textbf{ B 653} (2007) 60,
  \href{http://dx.doi.org/10.1016/j.physletb.2007.04.031}{\doi{10.1016/j.physletb.2007.04.031}},
\href{http://www.arXiv.org/abs/hep-ph/0703129}{\texttt{ arXiv:hep-ph/0703129}}.

\bibitem{Gong:2008hk}
\hrefCMSnoop {} {B.~Gong and J.-X. Wang, ``{QCD corrections to polarization of
  \JPsi\ and $\Upsilon$ at Tevatron and LHC}'',} \textit{ Phys. Rev.} \textbf{
  D 78} (2008) 074011,
  \href{http://dx.doi.org/10.1103/PhysRevD.78.074011}{\doi{10.1103/PhysRevD.78.074011}},
\href{http://www.arXiv.org/abs/0805.2469}{\texttt{ arXiv:0805.2469}}.

\bibitem{PRL.111.122001}
\hrefCMSnoop {} {J.-P. Lansberg and H.-S. Shao, ``Production of
  $\JPsi$+$\eta_c$ versus $\JPsi$+$\JPsi$ at the LHC: Importance of Real
  $\alpha_s^5$ Corrections'',} \textit{ Phys. Rev. Lett.} \textbf{ 111} (2013)
  122001,
\href{http://dx.doi.org/10.1103/PhysRevLett.111.122001}{\doi{10.1103/PhysRevLett.111.122001}}.

\bibitem{Aaij:2012dz}
\hrefCMSnoop {} {{LHCb Collaboration}, R.~Aaij {et~al.}, ``{Observation of
  \JPsi -pair production in pp collisions at $\sqrt{s}$ = 7 TeV}'',} \textit{
  Phys. Lett.} \textbf{ B 707} (2012) 52,
  \href{http://dx.doi.org/10.1016/j.physletb.2011.12.015}{\doi{10.1016/j.physletb.2011.12.015}},
\href{http://www.arXiv.org/abs/1109.0963}{\texttt{ arXiv:1109.0963}}.

\bibitem{Berezhnoy:2012xq}
\hrefCMSnoop {} {A.~V. Berezhnoy, A.~K. Likhoded, A.~V. Luchinsky, and A.~A.
  Novoselov, ``{Double $c\bar{c}$ production at LHCb}'',} \textit{ Phys. Rev.}
  \textbf{ D 86} (2012) 034017,
  \href{http://dx.doi.org/10.1103/PhysRevD.86.034017}{\doi{10.1103/PhysRevD.86.034017}},
\href{http://www.arXiv.org/abs/1204.1058}{\texttt{ arXiv:1204.1058}}.

\bibitem{Maltoni:2004hv}
\hrefCMSnoop {} {F.~Maltoni and A.~D. Polosa, ``{Observation potential for
  $\eta_b$ at the Tevatron}'',} \textit{ Phys. Rev.} \textbf{ D 70} (2004)
  054014,
  \href{http://dx.doi.org/10.1103/PhysRevD.70.054014}{\doi{10.1103/PhysRevD.70.054014}},
\href{http://www.arXiv.org/abs/hep-ph/0405082}{\texttt{ arXiv:hep-ph/0405082}}.

\bibitem{Braaten:2000cm}
\hrefCMSnoop {} {E.~Braaten, S.~Fleming, and A.~K. Leibovich, ``{NRQCD analysis
  of bottomonium production at the Tevatron}'',} \textit{ Phys. Rev.} \textbf{
  D 63} (2001) 094006,
  \href{http://dx.doi.org/10.1103/PhysRevD.63.094006}{\doi{10.1103/PhysRevD.63.094006}},
\href{http://www.arXiv.org/abs/hep-ph/0008091}{\texttt{ arXiv:hep-ph/0008091}}.

\bibitem{Jia:2006rx}
\hrefCMSnoop {} {Y.~Jia, ``{Which hadronic decay modes are good for $\eta_b$
  searching: double \JPsi\ or something else?}'',} \textit{ Phys. Rev.}
  \textbf{ D 78} (2008) 054003,
  \href{http://dx.doi.org/10.1103/PhysRevD.78.054003}{\doi{10.1103/PhysRevD.78.054003}},
\href{http://www.arXiv.org/abs/hep-ph/0611130}{\texttt{ arXiv:hep-ph/0611130}}.

\bibitem{Dermisek:2005ar}
\hrefCMSnoop {} {R.~Dermisek and J.~F. Gunion, ``{Escaping the large fine
  tuning and little hierarchy problems in the next to minimal supersymmetric
  model and $h \to aa$ decays}'',} \textit{ Phys. Rev. Lett.} \textbf{ 95}
  (2005) 041801,
  \href{http://dx.doi.org/10.1103/PhysRevLett.95.041801}{\doi{10.1103/PhysRevLett.95.041801}},
\href{http://www.arXiv.org/abs/hep-ph/0502105}{\texttt{ arXiv:hep-ph/0502105}}.

\bibitem{Domingo:2009tb}
\hrefCMSnoop {} {F.~Domingo, U.~Ellwanger, and M.-A. Sanchis-Lozano,
  ``{Bottomoniom spectroscopy with mixing of $\eta_b$ states and a light CP-odd
  Higgs}'',} \textit{ Phys. Rev. Lett.} \textbf{ 103} (2009) 111802,
  \href{http://dx.doi.org/10.1103/PhysRevLett.103.111802}{\doi{10.1103/PhysRevLett.103.111802}},
\href{http://www.arXiv.org/abs/0907.0348}{\texttt{ arXiv:0907.0348}}.

\bibitem{Domingo:2010am}
\hrefCMSnoop {} {F.~Domingo, ``{Updated constraints from radiative $\Upsilon$
  decays on a light CP-odd Higgs}'',} \textit{ JHEP} \textbf{ 04} (2011) 016,
  \href{http://dx.doi.org/10.1007/JHEP04(2011)016}{\doi{10.1007/JHEP04(2011)016}},
\href{http://www.arXiv.org/abs/1010.4701}{\texttt{ arXiv:1010.4701}}.

\bibitem{Aubert:2009as}
\hrefCMSnoop {} {{BaBar Collaboration}, B.~Aubert {et~al.}, ``{Evidence for the
  $\eta_b$(1S) Meson in Radiative $\Upsilon$(2S) Decay}'',} \textit{ Phys. Rev.
  Lett.} \textbf{ 103} (2009) 161801,
  \href{http://dx.doi.org/10.1103/PhysRevLett.103.161801}{\doi{10.1103/PhysRevLett.103.161801}},
\href{http://www.arXiv.org/abs/0903.1124}{\texttt{ arXiv:0903.1124}}.

\bibitem{babarhiggs}
\hrefCMSnoop {} {{BaBar Collaboration}, B.~Aubert {et~al.}, ``{Search for
  Dimuon Decays of a Light Scalar Boson in Radiative Transitions $\Upsilon \to
  \gamma A^0$}'',} \textit{ Phys. Rev. Lett.} \textbf{ 103} (2009) 081803,
  \href{http://dx.doi.org/10.1103/PhysRevLett.103.081803}{\doi{10.1103/PhysRevLett.103.081803}},
\href{http://www.arXiv.org/abs/0905.4539}{\texttt{ arXiv:0905.4539}}.

\bibitem{Chatrchyan:2012am}
\hrefCMSnoop {} {{CMS Collaboration}, S.~Chatrchyan {et~al.}, ``{Search for a
  light pseudoscalar Higgs boson in the dimuon decay channel in pp collisions
  at $\sqrt{s}=7$ TeV}'',} \textit{ Phys. Rev. Lett.} \textbf{ 109} (2012)
  121801,
  \href{http://dx.doi.org/10.1103/PhysRevLett.109.121801}{\doi{10.1103/PhysRevLett.109.121801}},
\href{http://www.arXiv.org/abs/1206.6326}{\texttt{ arXiv:1206.6326}}.

\bibitem{Sjostrand:2006za}
\hrefCMSnoop {} {T.~Sj{\"o}strand, S.~Mrenna, and P.~Z. Skands, ``{PYTHIA} 6.4
  physics and manual'',} \textit{ JHEP} \textbf{ 05} (2006) 026,
  \href{http://dx.doi.org/10.1088/1126-6708/2006/05/026}{\doi{10.1088/1126-6708/2006/05/026}},
\href{http://www.arXiv.org/abs/hep-ph/0603175}{\texttt{ arXiv:hep-ph/0603175}}.

\bibitem{Sjostrand:2007gs}
\hrefCMSnoop {} {T.~Sj{\"o}strand, S.~Mrenna, and P.~Z. Skands, ``{A Brief
  Introduction to PYTHIA 8.1}'',} \textit{ Comput. Phys. Commun.} \textbf{ 178}
  (2008) 852,
  \href{http://dx.doi.org/10.1016/j.cpc.2008.01.036}{\doi{10.1016/j.cpc.2008.01.036}},
\href{http://www.arXiv.org/abs/0710.3820}{\texttt{ arXiv:0710.3820}}.

\bibitem{Chatrchyan:2008zzk}
\hrefCMSnoop {} {{CMS Collaboration}, S.~Chatrchyan {et~al.}, ``The {CMS}
  experiment at the {CERN} {LHC}'',} \textit{ JINST} \textbf{ 3} (2008) S08004,
  \href{http://dx.doi.org/10.1088/1748-0221/3/08/S08004}{\doi{10.1088/1748-0221/3/08/S08004}}.

\bibitem{Agostinelli2003250}
\hrefCMSnoop {} {{GEANT 4 Collaboration}, S.~Agostinelli {et~al.}, ``Geant4 --
  A Simulation Toolkit'',} \textit{ Nucl. Instrum. Meth.} \textbf{ A 506}
  (2003) 250,
  \href{http://dx.doi.org/10.1016/S0168-9002(03)01368-8}{\doi{10.1016/S0168-9002(03)01368-8}}.

\bibitem{Fruhwirth:1987fm}
\hrefCMSnoop {} {R.~Fr{\"u}hwirth, ``{Application of Kalman filtering to track
  and vertex fitting}'',} \textit{ Nucl. Instrum. Meth.} \textbf{ A 262} (1987)
  444,
\href{http://dx.doi.org/10.1016/0168-9002(87)90887-4}{\doi{10.1016/0168-9002(87)90887-4}}.

\bibitem{PDG2012}
\hrefCMSnoop {} {{Particle Data Group}, J.~Beringer {et~al.}, ``{Review of
  Particle Physics (RPP)}'',} \textit{ Phys. Rev.} \textbf{ D 86} (2012)
  010001,
\href{http://dx.doi.org/10.1103/PhysRevD.86.010001}{\doi{10.1103/PhysRevD.86.010001}}.

\bibitem{CMS-PAS-SMP-12-008}
\href {http://cdsweb.cern.ch/record/1434360} {{CMS Collaboration}, ``Absolute
  Calibration of the Luminosity Measurement at {CMS}: {W}inter 2012 Update'',}
  CMS Physics Analysis Summary CMS-PAS-SMP-12-008, 2012.

\bibitem{sPlot}
\hrefCMSnoop {} {M.~{Pivk} and F.~R. {Le Diberder}, ``{sPlot}: {A} statistical
  tool to unfold data distributions'',} \textit{ Nucl. Instrum. Meth.} \textbf{
  A 555} (2005) 356,
  \href{http://dx.doi.org/10.1016/j.nima.2005.08.106}{\doi{10.1016/j.nima.2005.08.106}},
  \href{http://www.arXiv.org/abs/physics/0402083}{\texttt{
  arXiv:physics/0402083}}.

\bibitem{roofit}
\href {http://www.slac.stanford.edu/econf/C0303241/proc/papers/MOLT007.PDF}
  {W.~Verkerke and D.~P. Kirkby, ``{The RooFit toolkit for data modeling}'',}
  in \textit{ Computing in High-Energy and Nuclear Physics (CHEP 03)}.
\newblock 2003.
\newblock \href{http://www.arXiv.org/abs/physics/0306116}{\texttt{
  arXiv:physics/0306116}}.
\newblock
eConf C~0303241 MOLT007.

\bibitem{CMS-PAS-TRK-10-002}
\href {http://cdsweb.cern.ch/record/1279139} {{CMS Collaboration},
  ``Measurement of Tracking Efficiency'',} CMS Physics Analysis Summary
  CMS-PAS-TRK-10-002, 2010.

\bibitem{MUO-10-004}
\hrefCMSnoop {} {{CMS Collaboration}, S.~Chatrchyan {et~al.}, ``{Performance of
  CMS muon reconstruction in pp collision events at $\sqrt{s} = 7$ TeV}'',}
  \textit{ JINST} \textbf{ 7} (2012) P10002,
  \href{http://dx.doi.org/10.1088/1748-0221/7/10/P10002}{\doi{10.1088/1748-0221/7/10/P10002}},
  \href{http://www.arXiv.org/abs/1206.4071}{\texttt{ arXiv:1206.4071}}.

\bibitem{ref:crystalball}
\href {http://www.slac.stanford.edu/pubs/slacreports/slac-r-236.html} {M.~J.
  Oreglia, ``{A Study of the Reactions $\psi^\prime \to \gamma \gamma
  \psi$}'',} {Ph.D. Thesis}, Stanford University, 1980.
\newblock SLAC-R-236.

\bibitem{Chao:2012iv}
K.-T. Chao\hrefCMSnoop {} { {et~al.}, ``{\JPsi\ polarization at Hadron
  Colliders in Nonrelativistic QCD}'',} \textit{ Phys. Rev. Lett.} \textbf{
  108} (2012) 242004,
  \href{http://dx.doi.org/10.1103/PhysRevLett.108.242004}{\doi{10.1103/PhysRevLett.108.242004}},
\href{http://www.arXiv.org/abs/1201.2675}{\texttt{ arXiv:1201.2675}}.

\end{thebibliography}\endgroup

\cleardoublepage \appendix\section{The CMS Collaboration \label{app:collab}}\begin{sloppypar}\hyphenpenalty=5000\widowpenalty=500\clubpenalty=5000\textbf{Yerevan Physics Institute,  Yerevan,  Armenia}\\*[0pt]
V.~Khachatryan, A.M.~Sirunyan, A.~Tumasyan
\vskip\cmsinstskip
\textbf{Institut f\"{u}r Hochenergiephysik der OeAW,  Wien,  Austria}\\*[0pt]
W.~Adam, T.~Bergauer, M.~Dragicevic, J.~Er\"{o}, C.~Fabjan\cmsAuthorMark{1}, M.~Friedl, R.~Fr\"{u}hwirth\cmsAuthorMark{1}, V.M.~Ghete, C.~Hartl, N.~H\"{o}rmann, J.~Hrubec, M.~Jeitler\cmsAuthorMark{1}, W.~Kiesenhofer, V.~Kn\"{u}nz, M.~Krammer\cmsAuthorMark{1}, I.~Kr\"{a}tschmer, D.~Liko, I.~Mikulec, D.~Rabady\cmsAuthorMark{2}, B.~Rahbaran, H.~Rohringer, R.~Sch\"{o}fbeck, J.~Strauss, A.~Taurok, W.~Treberer-Treberspurg, W.~Waltenberger, C.-E.~Wulz\cmsAuthorMark{1}
% ...
\vskip\cmsinstskip
% ...
1:~~Also at Vienna University of Technology, Vienna, Austria\\
2:~~Also at CERN, European Organization for Nuclear Research, Geneva, Switzerland\\
% ...

\end{sloppypar}
\end{document}